\documentclass[preprint,pdf]{iucr}              

\usepackage{graphicx}
\usepackage{amssymb}
\usepackage{marvosym}
\usepackage{upgreek}
\usepackage{siunitx}
\usepackage{comment}
\usepackage{amsmath}

\begin{document}                  



\title{The new neutron grating interferometer at the ANTARES beamline \\- Design, Principle, and Applications -}


\cauthor[a,b]{Tommy}{Reimann}{tommy.reimann@frm2.tum.de}{}
\author[a]{Sebastian}{M\"uhlbauer}
\author[c]{Michael}{Horisberger}
\author[b]{Peter}{B\"oni}
\author[a,b]{Michael}{Schulz}

\aff[a]{Heinz Maier-Leibnitz Zentrum (MLZ), Technische Universit\"at M\"unchen, Lichtenbergstr. 1, 85748 Garching, \country{Germany}}
\aff[b]{Physik-Department E21, Technische Universit\"at M\"unchen, James-Franck-Str. 1, 85748 Garching, \country{Germany}}
\aff[c]{Neutron Optics and Scientific Computing Group, Paul Scherrer Institut, 5232 Villigen PSI, \country{Switzerland}}









\maketitle                        


\begin{abstract}
Neutron grating interferometry is an advanced method in neutron imaging that allows the simultaneous recording of the transmission, the differential phase and the dark-field image. Especially the latter has recently received high interest because of its unique contrast mechanism which marks ultra-small-angle neutron scattering within the sample. Hence, in neutron grating interferometry, an imaging contrast is generated by scattering of neutrons off micrometer-sized inhomogeneities. Although the scatterer cannot be resolved it leads to a measurable local decoherence of the beam. Here, a report is given on the design considerations, principles and applications of a new neutron grating interferometer which has recently been implemented at the ANTARES beamline at the Heinz Maier-Leibnitz Zentrum. Its highly flexible design allows to perform experiments such as directional and quantitative dark-field imaging which provide spatially resolved information on the anisotropy and shape of the microstructure of the sample. A comprehensive overview of the nGI principle is given, followed by theoretical considerations to optimize the setup performance for different applications. Furthermore, an extensive characterization of the setup is presented and its abilities are demonstrated on selected case studies: (i) dark-field imaging for material differentiation, (ii) directional dark-field imaging to mark and quantify micrometer anisotropies within the sample and (iii) quantitative dark-field imaging, providing additional size information on the sample's microstructure by probing its autocorrelation function. 
\end{abstract}


\section{Introduction}

Neutron radiography is a nondestructive imaging technique, which provides information about the interior of an object with high spatial resolution by using neutron radiation \cite{anderson2009neutron}. In contrast to x-ray radiography, neutron imaging is sensitive to some light elements as hydrogen or lithium, while most heavy elements as e.g. lead and aluminum can easily be penetrated. Consequently, this method is routinely applied in fields such as cultural heritage research \cite{Mannes:2014-03-01T00:00:00:1354-2575:137}, materials science \cite{Kardjilov2011248}, engineering \cite{Gruenzweig_chain_saw}, and geology \cite{Hess01122011}, whenever x-rays fail to generate sufficient imaging contrast or lack of penetration.
\\Nowadays, spatial resolutions down to \SI{50}{\micro\meter} are routinely obtained by means of neutron imaging which are limited by the geometric resolution of the beamline (L/D-ratio) and the resolution obtainable with neutron detectors \cite{anderson2009neutron}. Several approaches have been proposed to investigate smaller structures. They are either based on the direct magnification of the image by focusing neutron optics \cite{Lui2013} or on the improvement of the detector resolution \cite{Trtik2015169}. In the following, we will concentrate on a third approach which is provided by neutron grating interferometry (nGI) \cite{Grunzweig_Rev_Sci_Inst_2008}. 
\\nGI is an advanced neutron imaging method which allows the simultaneous recording of the neutron transmission image (TI), differential phase contrast image (DPC), and the dark-field image (DFI) \cite{Grunzweig_Rev_Sci_Inst_2008}\cite{:/content/aip/journal/apl/93/11/10.1063/1.2975848}. A nGI setup consists of two neutron absorption gratings and one neutron phase grating implemented in a neutron imaging beamline. While the spatial resolution of nGI is limited by the same restrictions as discussed above, the contrast of the DFI is generated by ultra small-angle neutron scattering (USANS) off micrometer sized structures within the sample \cite{PhysRevLett.101.123902}. Hence, DFI marks the presence of micrometer sized inhomogeneities of the nuclear and magnetic scattering length density by their scattering signature, though they cannot be resolved directly. In this way, the DFI is sensitive to magnetic scattering, without the need for polarization analysis \cite{Kardjilov2008}. Consequently, nGI has been applied not only for the differentiation \cite{PhysRevLett.101.123902} and testing \cite{:/content/aip/journal/jap/107/3/10.1063/1.3298440} of materials, but also for the investigation of magnetic micrometer structures in ferromagnets \cite{PhysRevLett.101.025504}\cite{:/content/aip/journal/apl/93/11/10.1063/1.2975848} or superconductors \cite{Reimann2015a}\cite{Reimann2015b}.
\\The improving theoretical understanding of the nGI contrast mechanism has recently triggered the transition of nGI towards a quantitative method providing detailed information about the microstructure of the sample \cite{Strobl2014}\cite{Lynch2011}: Details of its morphology can be obtained by wavelength dependent dark-field imaging (quantitative dark-field imaging \cite{Betz2015}). Moreover, anisotropies and textures on the $\upmu$m scale can be detected by analyzing the DFI variation during a rotation of the grating setup around the beam axis (directional dark-field imaging \cite{PhysRevB.82.214103}). 
\\Here we report about the setup and applications of a new nGI which has recently been implemented at the ANTARES imaging beamline at the Heinz Maier-Leibnitz Zentrum (MLZ). The setup was designed to perform directional and quantitative dark-field imaging even in combination with complex sample environments. Its main advantages over existing nGIs are (i) the high neutron flux available at ANTARES, (ii) the capability to rotate all gratings simultaneously around the beam axis, (iii) the ability to flexibly adjust the neutron spectrum by using different filters or monochromators and (iv) its design which allows a combination with the various sample environments (e.g. non-ambient temperatures and magnetic fields) specifically build for neutron imaging. 
\\This paper is structured as follows: In Sec. \ref{Sec_nGI_principle} we give an overview about the nGI imaging principle, followed by a theoretical discussion of the DFI contrast modality. In this context, we present a simple model which allows to estimate the visibility for arbitrary neutron wavelength $\lambda$, setup distances and wavelength distributions which can be employed to tune the sensitivity of the setup. Sec. \ref{Sec_nGI_ANTARES} describes the ANTARES instrument, its nGI setup and the grating manufacturing in detail. The characterization of the nGI setup, regarding visibility, flux and the sensitivity of the DFI contrast modality to structures of different size is given in Sec. \ref{Sec_char}. Moreover, the potential of a quantitative DFI evaluation using ANTARES is demonstrated. Finally, further typical nGI applications are presented in the following section, highlighting its unique capabilities such as dark-field imaging for the differentiation of materials having similar transmission (Sec. \ref{Mat_diff}) and directional dark-field imaging to extract microstructural anisotropies (Sec. \ref{Directional_DFI}).

\section{The principles of neutron grating interferometry}
\label{Sec_nGI_principle}

\subsection{The imaging principle of nGI}
\begin{figure}\centering%
\includegraphics[width=0.5\textwidth]{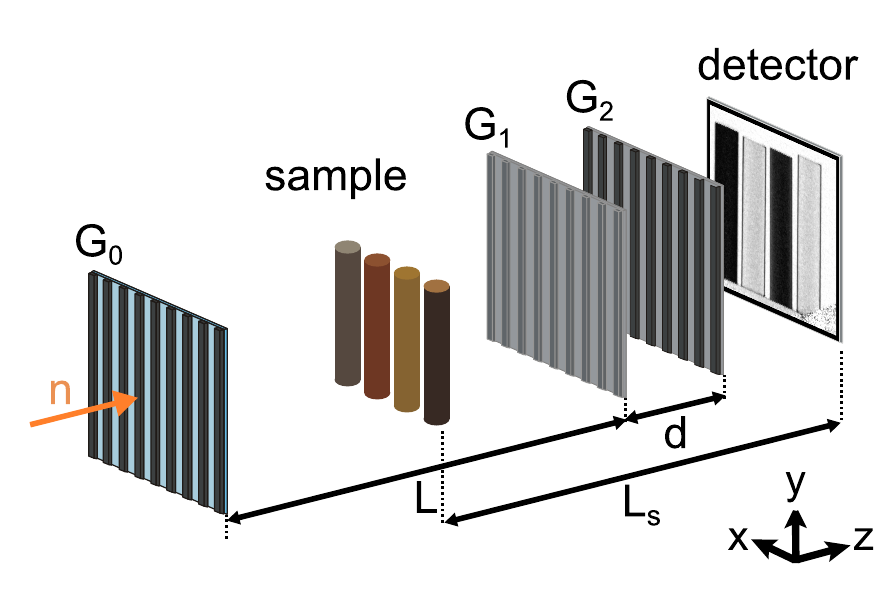}%
\caption{Illustration of the nGI setup. The setup consists of the source grating G$_0$, the phase grating G$_1$, the analyzer grating G$_2$, and a neutron imaging detector. The sample may be placed either between G$_0$ and G$_1$ or G$_1$ and G$_2$.}%
\label{Setup_schematic}%
\end{figure}
\begin{figure}\centering%
\includegraphics[width=0.5\textwidth]{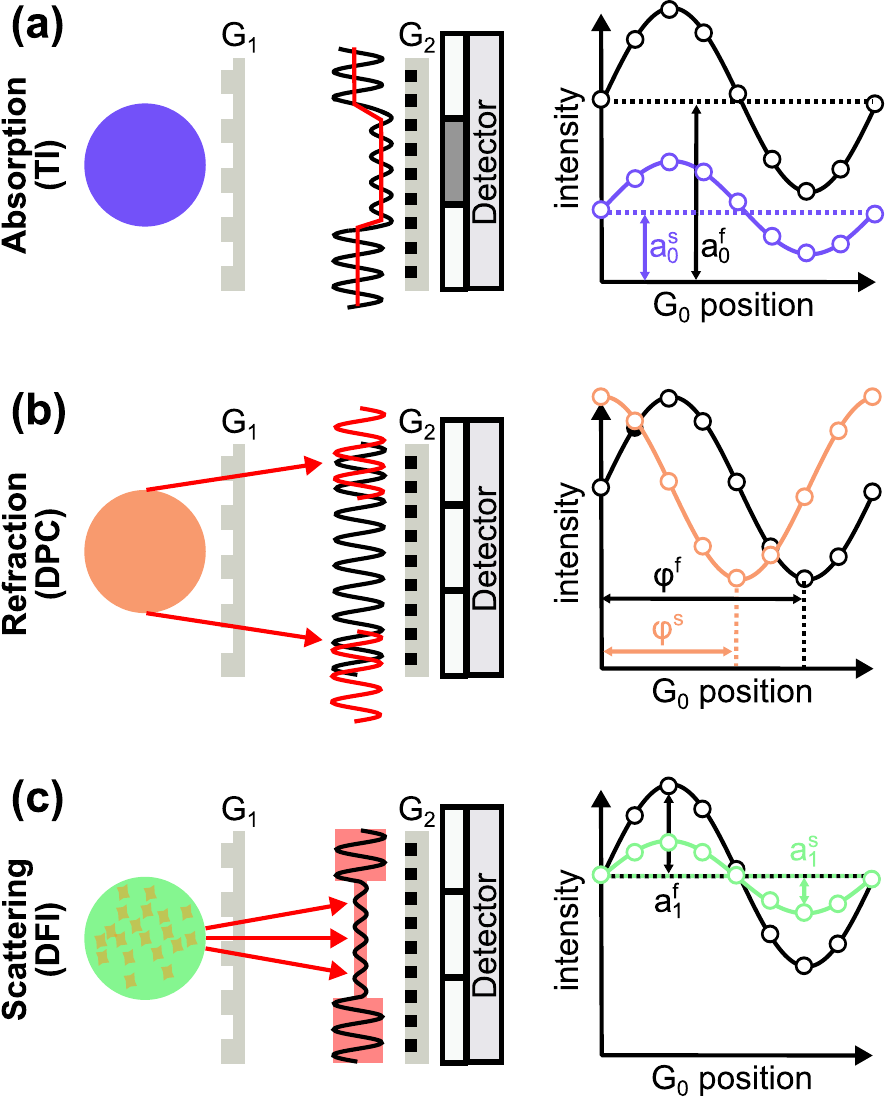}%
\caption{The nGI contrast mechanism: Illustration how absorption (a), refraction (b) and scattering (c) influence the intensity oscillation $I(x_{\mathrm{g}},j,l)$ and hence generate contrast in the TI, DPC and DFI contrast channels, respectively.}%
\label{Contrast_mechanism}%
\end{figure}

A neutron grating interferometer is a realization of a Talbot-Lau interferometer for neutrons \cite{Lau1948} that is implemented into a neutron imaging beamline. A schematic depiction of its main components is shown in Fig. \ref{Setup_schematic}. The absorption grating G$_0$ (periodicity $p_0 \sim \SI{}{\milli\meter}$), which is located shortly behind the neutron pinhole, generates an array of coherent, but mutually incoherent line sources. A distance $L$ ($\sim\SI{}{\meter}$) downstream, the phase grating G$_1$ ($p_1\sim\SI{}{\micro\meter}$) imprints a periodic phase modulation onto the neutron wave front. Due to the Talbot effect \cite{Lau1948}, this phase modulation generates an intensity modulation behind the grating \cite{Gruenzweig_thesis}, often called "Talbot carpet", which is maximal at the fractional Talbot distances $d_\mathrm{n}$ having an odd $n$ but vanishes at even $n$ \cite{Hipp:14}:
\begin{equation}
d_{\mathrm{n}}=\frac{n}{16}d_{\mathrm{T}}=n\frac{p_1^2}{8\lambda}
\label{Talbot-distance}
\end{equation}
\\The generated interference pattern has approximately half the periodicity of the phase grating. Hence, it is not directly accessibly by an imaging detector as its pitch is well below the detector resolution. Therefore, the analyzer grating $G_2$ ($p_2\approx p_1/2$) is introduced at a distance $d\sim\SI{}{\centi\meter}$ from G$_1$, directly in front of the detector \cite{PhysRevLett.96.215505}. As the periodicities of the interference pattern and G$_2$ match, the transmitted intensity $I$ is minimized if the grating lines cover the interference maxima and vice versa. Hence, a translation $x_{\mathrm{g}}$ of one of the gratings G$_{\mathrm{i}}$ ($i=1,2,3$) perpendicular to the beam and to the grating lines will result in an intensity oscillation in each detector pixel ($j$,$l$) which can be approximated by:
\begin{equation}
I(x_{\mathrm{g}},j,l) = a_0(j,l)+a_1(j,l)\cos\left(\frac{2\pi x_{\mathrm{g}}}{p_{\mathrm{i}}}-\varphi(j,l)\right)
\end{equation}
$a_0$, $a_1$ and $\varphi$ are the offset, amplitude and phase of the oscillation, respectively \cite{Grunzweig_Rev_Sci_Inst_2008}. 
\\An nGI scan measures the change of $I(x_{\mathrm{g}},j,l)$ due to the influence of the sample onto the interference pattern. By recording neutron images at different G$_0$-positions $x_{\mathrm{g}}$, a determination of $a_0$, $a_1$ and $\varphi$ is possible via a least square fit \cite{:/content/aip/journal/rsi/85/1/10.1063/1.4861199} or a Fast Fourier Transformation of the data \cite{Chabior2011a}. If this stepping scan is performed once without (f) and once with a sample (s) inserted in the interferometer, the TI, DPC, and DFI can be calculated from the dataset \cite{:/content/aip/journal/apl/93/11/10.1063/1.2975848}\cite{PhysRevLett.96.215505}:
\begin{align}
TI(j,l) &= \frac{a^{\mathrm{s}}_0(j,l)}{a^{\mathrm{f}}_0(j,l)}\\
DPC(j,l) &= \varphi^{\mathrm{s}}(j,l)-\varphi^{\mathrm{f}}(j,l)\\
DFI(j,l) &=\frac{a^{\mathrm{s}}_1(j,l)a^{\mathrm{f}}_0(j,l)}{a^{\mathrm{s}}_0(j,l)a^{\mathrm{f}}_1(j,l)}\label{Equ_DFI_Gen}
\end{align}
A summarizing overview, illustrating how the different nGI contrast channels are influenced is given in Fig. \ref{Contrast_mechanism}: Neutron absorption leads to an attenuation of the whole interference pattern and is seen in the TI only. In contrast, refraction leads to a deflection of the neutron beam and hence, to a phase shift of the intensity oscillation recorded by the DPC. Finally, the DFI is influenced by neutron scattering under ultra-small angles as this does not change the average intensity reaching a detector pixel, but smears the interference pattern \cite{PhysRevLett.101.123902}. Consequently, an nGI scan provides three complementary physical information and clearly broadens the information range obtainable by neutron imaging \cite{Kardjilov2011248}. In the following, we will focus on the DFI contrast modality and explain its dependence on the microstructure of the sample. Contrary to the definition of an imaging contrast, we will use the following nomenclature: A decreasing DFI according to Equation \ref{Equ_DFI_Gen} corresponds to a decreasing DFI contrast.

\subsection{The dark-field contrast modality}
\label{Sec_QDFI}
In the following we will discuss the DFI contrast mechanism and how the DFI contrast is linked to the setup parameters and the microstructural properties of the sample. For x-ray grating interferometry, the DFI contrast has been treated by rigorous wave propagation calculations in Ref. \cite{Lynch2011} and \cite{Yashiro:10}. Their results demonstrate a connection between the DFI contrast and the autocorrelation function of the refraction index within the microstructure of the sample. A complementary, but more general theoretical approach has been given in Ref. \cite{Strobl2014}. In this work, it has been assumed that scattering within the sample, involving a momentum transfer $q_\perp$ perpendicular to the grating lines, leads to a fractional deflection of the neutron beam. Hence, the scattering is accompanied by a phase shift $\Delta\varphi=\xi_{\mathrm{GI}}q_\perp$ of a part of the interference pattern, generated by G$_1$. Consequently, the DFI reduction results from a superposition of the undisturbed and the deflected parts of interference pattern. The setup specific correlation length $\xi_{\mathrm{GI}}$ is defined as:
\begin{equation}
\xi_{\mathrm{GI}}=\frac{\lambda L_{\mathrm{s}}^{\mathrm{eff}}}{p_2}
\label{xiGI}
\end{equation} 
and depends on an effective sample-to-detector-distance $L_{\mathrm{s}}^{\mathrm{eff}}$. By inserting the distance $L$ between G$_0$ and G$_1$ and the sample-to-detector-distance $L_\mathrm{s}$ its value can be calculated according to \cite{:/content/aip/journal/jap/106/5/10.1063/1.3208052}:
\begin{equation}
L_{\mathrm{s}}^{\mathrm{eff}}=\begin{cases}
L_{\mathrm{s}}& \text{for $L_{\mathrm{s}}<d$}\\
(L+d-L_{\mathrm{s}})\frac{d}{L}&\text{for $L_{\mathrm{s}}>d$}\\
\end{cases}
\end{equation}
\\In the special case of isotropic scattering, the DFI contrast variation can be derived as:
\begin{equation}
DFI(\xi_{\mathrm{GI}})=\exp\left\{\Sigma t\left[G\left(\xi_{\mathrm{GI}}\right)-1\right]\right\}
\label{DFI_xi}
\end{equation}
in which $\Sigma$ is the macroscopic scattering cross-section of the material and $t$ the sample thickness. The real space correlation function $G$ is the cosine Fourier transform of the scattering function which is routinely measured in e.g. small-angle neutron scattering \cite{Andersson:aj5110} or ultra-small-angle neutron scattering \cite{Rehm:he5570}. A geometrical interpretation of $G$ can be found in Ref. \cite{Krouglovpe0091}. Note that these relations again show that the DFI contrast is caused by the microstructure of the sample, as $\xi_{\mathrm{GI}}$ amounts a few micrometers in a typical nGI setup.
\\Four distinct conclusions can be drawn from the derivation above: (i) The DFI contrast depends exponentially on the sample thickness in beam direction. Hence, tomographic DFI reconstructions can be based on the same algorithms as attenuation based tomography \cite{PhysRevLett.101.123902}. (ii) A variation of $\xi_{\mathrm{GI}}$ via a wavelength or distance scan allows to directly measure the real space correlation function of the material within the boundaries given by the accessible wavelength band or setup distances. Hence, the deduction of quantitative information is possible (Sec. \ref{QDFI}). (iii) As the dark-field signal is generated by the microstructure of the sample, a material specific dark-field extinction coefficient can be defined \cite{PhysRevB.88.125104} which differs from the attenuation coefficient and provides additional imaging contrast (see Sec. \ref{Mat_diff}). (iv) The DFI is insensitive to scattering contributions parallel to the grating lines. Consequently, if the scattering function and hence the underlying microstructure of an object are anisotropic, its DFI will depend on the rotation angle $\omega$ of the grating lines around the beam axis. Hence, detailed information about the microstructural orientation can be obtained by a DFI($\omega$) scan (Sec. \ref{Directional_DFI}). 

\subsection{The optimized setup geometry: Visibility and flux considerations}
\label{Sec_wavelength_V}
The signal-to-noise ratio of the DPC and DFI data depends on the visibility $V$ of the measured interference pattern, given by $ V =a_1(j,l)/a_0(j,l)$ \cite{Chabior2011a}. In order to maximize $V$, the periodicity of the gratings $p_0$, $p_1$ and $p_2$, their distances $L$ and $d$ as well as the neutron wavelength $\lambda$ have to satisfy several mutual relations. These necessary considerations are extensively described in e.g. \cite{Grunzweig_Rev_Sci_Inst_2008} and \cite{Chabior_thesis}. 
\\However, the aforementioned relations optimize the visibility only for a mono\-chromatic nGI measurement under the assumption of perfectly absorbing gratings. The neutron flux and a residual transmission of the grating lines which also influence the image quality are not considered. Furthermore, some nGI measurements even require a deviation from these optimal setup parameters: In particular, time-resolved measurements could benefit from a reduction of the setup length, as this enhances the neutron flux at the sample position which is proportional to $(L+d)^{-2}$. Moreover, for quantitative DFI measurements (Sec. \ref{potential_QDFI}), a variation of the neutron wavelength and its wavelength distribution is necessary to vary the probed correlation length $\xi_\mathrm{GI}$. In the following, we will describe which of the relations given in Ref. \cite{Grunzweig_Rev_Sci_Inst_2008} necessarily have to be fulfilled. Furthermore, a simple model is developed that allows to estimate the visibility of the nGI for arbitrary setup configurations by determining only one fit parameter. Hence, the influence of (i) the distance $d$, (ii) the wavelength $\lambda$ and (iii) the wavelength distribution onto the visibility can be easily determined in order to optimize a setup according to the requirements of the experiment. Our model includes the geometry of the setup and the wavelength dependencies of the Talbot effect as well as of the grating absorption. 

\subsubsection{The geometric relation of $d$ and $L$:}
\label{subdl}
If the distance $L$ is altered, the setup parameters have to be adjusted, such that the interference patterns originating from different slits of G$_0$ are still constructively superimposed onto G$_2$. This is guaranteed as long as the theorem of intersecting lines:
\begin{equation}
p_0\frac{d}{L}=p_2
\label{Equ_Strahlenstatz}
\end{equation}
is fulfilled \cite{PhysRevLett.96.215505}. A deviation from this relation has to be avoided, as it will strongly decrease the visibility and the visibility will become dependent on the pinhole diameter\footnote{If relation \ref{Equ_Strahlenstatz} is not fulfilled, the interference patterns originating from different slits of G$_0$ will be superimposed, each revealing a phase shift of $p_0d(Lp_2)^{-1}$. Hence, if the pinhole diameter (the divergence of the beam) is increased and more slits are illuminated, the interference pattern will be smeared out accordingly.}. In order to prevent a mismatch of the interference pattern generated by G$_1$ with G$_2$, which would result in horizontal Moir$\acute{e}$ streaks at the detector \cite{Wang:11}, $d$ and $L$ have furthermore to be scaled according to:
\begin{equation}
\frac{d}{L}=(\frac{2p_2}{p_1}-1)=\mathrm{const.}
\label{Magnification}
\end{equation}   
Consequently, $p_0$ is not affected by a variation of $L$, but $d$ scales proportionally.  

\subsubsection{The influence of $d$ on the visibility: }
Besides the geometrical considerations discussed above, the shape of the interference pattern (Talbot carpet) behind G$_1$ has to be considered \cite{Hipp:14}. A reduced setup length will result in a diminished visibility, because $d$ is driven out of the first fractal Talbot distance $d_1=p_1^2/8\lambda$ to comply with Equ. \ref{Equ_Strahlenstatz}. For x-rays, this problem was numerically treated in Ref. \cite{Hipp:14} in which the visibility of a perfectly coherent grating interferometer was calculated as function of the reduced propagation distance $\eta=\frac{d}{d_{\mathrm{T}}}\propto\frac{L}{d_{\mathrm{T}}}$ and the phase shift $\phi$ introduced by G$_1$. While not explicitly stated in Ref. \cite{Hipp:14}, their data clearly suggests that the visibility $V_{\mathrm{T}}$ of a perfect setup is well described by:
\begin{equation}
V_{\mathrm{T}}(\phi,\eta)=\frac{1}{2}\left(1-\cos\phi\right)\left|\sin(8\pi\eta)\right|
\label{V_T}
\end{equation}
Hence, a variation of the setup length leads to a sinusoidal reduction of the visibility . 

\subsubsection{Wavelength dependence of the visibility: }
It seems reasonable to compensate the reduction of visibility due to the changed setup length by an adjustment of $d$ and the fractional Talbot distance $d_1$ via a variation of the wavelength. However, a change of $\lambda$ influences not only the Talbot distance, but also the neutron phase shift introduced by G$_1$.  Furthermore, caused by the limited thickness of the Gd absorption layers, the visibility is also affected by the wavelength dependent transmission of the gratings G$_0$ and G$_2$ \cite{Chabior201271}. Hence, these contributions have to be considered separately:
\\Both $\eta$ as well as $\phi$ depend linearly on the neutron wavelength \cite{Grunzweig_Rev_Sci_Inst_2008}:
\begin{align}
\phi &=n_{\mathrm{sdl}}h_1\lambda\label{phi}\\
\eta &=\frac{d}{2p_1^2}\lambda\label{eta}
\end{align}
in which $n_{\mathrm{sdl}}=\SI{2.079e14}{\per\meter}$ is the scattering length density of the G$_1$ material Si \cite{doi:10.1080/10448639208218770} and $h_\mathrm{i}$ is the height of the grating lines of G$_\mathrm{i}$. Hence, the wavelength dependence of the visibility can be evaluated by combining Equ. \ref{V_T},\ref{phi} and \ref{eta} to yield:
\begin{equation}
V_{\mathrm{T}}(\lambda,d)=\frac{1}{2}\left[1-\cos\left(n_{\mathrm{sdl}}h_1\lambda\right)\right]\left|\sin\left(\frac{4\pi d}{p_1^2}\lambda\right)\right|
\end{equation}
\begin{figure}\centering%
\includegraphics{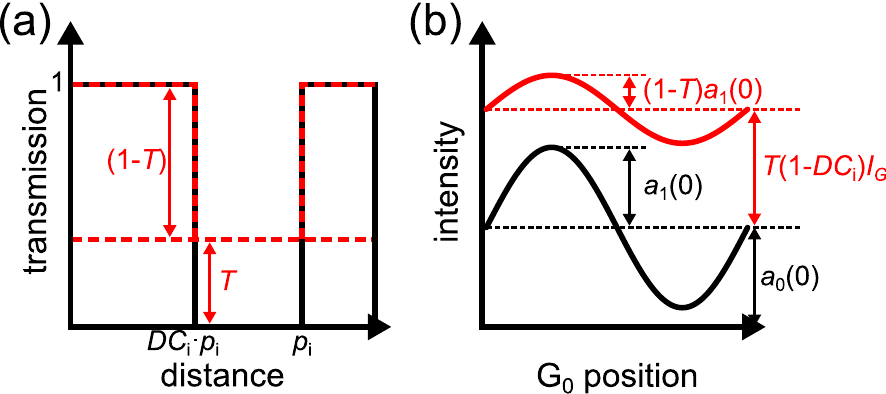}%
\caption{Reduction of visibility caused by finite transmission through the grating lines: (a) Simplified transmission profile of the absorption gratings. (b) If the transmission through the grating lines $T$ is non-zero the intensity oscillation at the detector is reduced and superimposed by a constant background.}
\label{Fig_Profile}%
\end{figure}
In addition, the effect of the finite grating line transmission on the visibility can be estimated by the following considerations: The absorption gratings G$_\mathrm{i}$ (i=0,2) having periodicity $p_{\mathrm{i}}$ and a duty cycle of $DC_{\mathrm{i}}$ will have a transmission profile as illustrated in Fig. \ref{Fig_Profile} a. In the case of a perfect grating, the transmission through the slits is unity, whereas the transmission of the grating lines, denoted as $T$, is zero. Therefore, the visibility will have an initial value $V_{\mathrm{Gi}}(T=0)=a_1(T=0)/a_0(T=0)$, in which $a_0(0)$ is given as $DC_{\mathrm{i}}$ multiplied by the neutron intensity incident on the grating $I_{\mathrm{G}}$. However, if $T$ is non-zero, the transmitted intensity will increase according to:
\begin{equation}
a_0(T)=\left[DC_{\mathrm{i}}+T(1-DC_{\mathrm{i}})\right]I_{\mathrm{G}}
\label{a0x}
\end{equation}
On the other hand, the intensity oscillation measured at the detector will correspond to the initial one, reduced by a factor of $(1-T)$, which is superimposed by a constant offset given as $T\cdot I_{\mathrm{G}}$ (compare red curve in Fig. \ref{Fig_Profile}b). Hence $a_1(T)$ scales as $(1-T)$:
\begin{equation}
a_1(T)=(1-T)a_1(0)=(1-T)V_{\mathrm{Gi}}(0)DC_{\mathrm{i}}I_{\mathrm{G}}
\label{a1x}
\end{equation}
For low neutron energies, the wavelength dependence of the neutron transmission $T$ can be approximated by:
\begin{equation}
T(\lambda)=\exp\left[-h_{\mathrm{i}}\sigma\frac{\lambda}{\SI{1.8}{\angstrom}}\right]
\label{xvonlamb}
\end{equation}
where $\sigma = \SI{1502.645}{\per\cm}$ is the macroscopic absorption cross section of Gd at $\SI{1.8}{\angstrom}$ \cite{doi:10.1080/10448639208218770}. By combining Equations \ref{a0x},\ref{a1x} and \ref{xvonlamb}, the wavelength dependence of the visibility caused by the finite transmission through the grating lines is derived as:

\begin{equation}
\frac{V_{\mathrm{Gi}}(T(\lambda))}{V_{\mathrm{Gi}}(0)}=\frac{1}{{V_{\mathrm{Gi}}(0)}}\frac{a_1(T(\lambda))}{a_0(T(\lambda))}=\frac{DC_{\mathrm{i}}}{DC_{\mathrm{i}}+\left[\exp(h_{\mathrm{i}}\sigma \frac{\lambda}{\SI{1.8}{\angstrom}})-1\right]^{-1}}
\end{equation}
At last, the wavelength dependence of the visibility is the product of the contributions from G$_0$, G$_1$ and the Talbot carpet:
\begin{equation}
V(\lambda,d)=V_0V_{\mathrm{T}}(\lambda,d)V_{\mathrm{G0}}(\lambda)V_{\mathrm{G2}}(\lambda)
\label{End_visibility}
\end{equation}
Hence, the complex mutual dependence of the different visibility contributions is reduced to an equation of only one unknown parameter $V_0$ which can be easily deduced experimentally. $V_0$ can be interpreted as maximal achievable visibility of the setup which depends on the quality of the gratings \cite{Chabior201271}, the coherence of the beam hitting G$_1$ \cite{PhysRevLett.94.164801}, and the smearing of the interference pattern due to the divergence of the beam \cite{Gruenzweig_thesis}. 

\subsubsection{The effect of a wavelength distribution onto the visibility: }

Finally, caused by the limitation of flux in neutron imaging, the neutron spectrum is not purely monochromatic in an nGI measurement. Hence, the influence of a wavelength distribution $f(\lambda)$ on the visibility has to be considered as well. It can be quantified according to:
\begin{equation}
V=A\int f(\tilde{\lambda})V(\tilde{\lambda})d\tilde{\lambda}
\label{Vis_distribution}
\end{equation}
in which A is defined by the normalization condition $A^{-1}=\int f(\tilde{\lambda})d\tilde{\lambda}$ \cite{Hipp:14}. nGI is typically used in combination with a neutron velocity selector (NVS) whose wavelength spread can be approximated by a triangular distribution \cite{FRIEDRICH1989547}. In this case, Equ. \ref{Vis_distribution} reduces to:
\begin{equation}
V_{\mathrm{NVS}}(\lambda,\Delta\lambda)=\frac{1}{\Delta\lambda}\int^{\lambda+\Delta\lambda}_{\lambda-\Delta\lambda}V(\tilde{\lambda})\left[1+\left(2\Theta(\tilde{\lambda}-\lambda)-1\right)\left(-\frac{\tilde{\lambda}}{\Delta\lambda}+\frac{\lambda}{\Delta\lambda}\right)\right]d\tilde{\lambda}
\label{NVS_distribution}
\end{equation}
Here, $\Theta(\tilde{\lambda}-\lambda)$ is the Heaviside step function. Based on the presented formulas, the setup visibility can be estimated for arbitrary configurations (wavelength, neutron filters, geometry) and be adjusted to the specific problem under investigation.

\section{The neutron grating interferometer at ANTARES}
\label{Sec_nGI_ANTARES}
\subsection{The ANTARES beamline}
\begin{figure}\centering%
\includegraphics[width=0.75\textwidth]{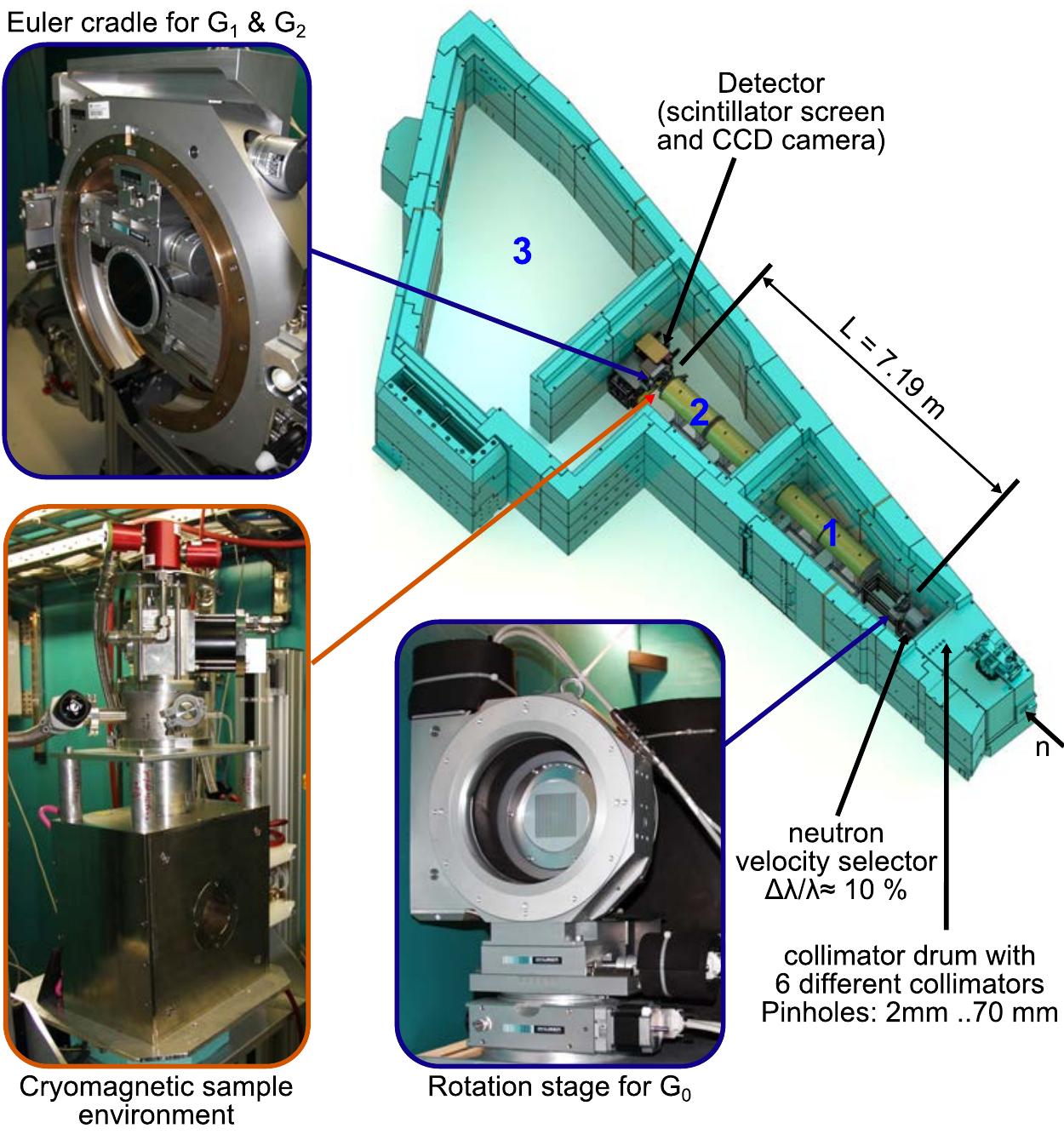}%
\caption{The nGI-setup at ANTARES: Drawing of the ANTARES beamline showing the main components of the nGI. For details see text.}%
\label{Antares-Setup}%
\end{figure}

ANTARES is a multi-purpose imaging beamline which is located at the beam port 4a of the FRM II reactor \cite{Calzada200950}. It provides a mixed spectrum of cold and thermal neutrons, peaked at $\lambda = \SI{1.6}{\angstrom}$ \cite{Tremsin2015}. The pinhole diameter can be varied between 2 and \SI{36}{\milli\meter} to adjust the geometrical resolution (L/D-ratio) of the instrument. The beamline is separated in three chambers (see Fig. \ref{Antares-Setup}). The first chamber (1) contains various beam shaping devices: a neutron velocity selector (Astrium NVS) providing a minimum wavelength of \SI{2.95}{\angstrom} with a wavelength spread $\Delta\lambda / \lambda$ of 10 \%, a double crystal monochromator (\SI{1.4}{\angstrom} -\SI{6.0}{\angstrom}, $\Delta\lambda / \lambda$ $\approx$ 3 \%), and a neutron filter wheel which includes a bismuth, a lead, a sapphire and a beryllium filter. 
\\Helium filled flight tubes transport the neutron beam to the remaining chambers (2 \& 3). Each of them contains a sample position equipped with a neutron imaging detector which is composed of a ${}^6$LiF/ZnS scintillator and a CCD camera (Andor IkonL 4Mpix). The maximum achievable neutron flux amounts to \SI{1d8}{\centi\meter^{-2}\second^{-1}} at $\mathrm{L/D}= 400$. The ANTARES instrument can be routinely used for neutron radiography and tomography, Bragg edge imaging \cite{ADMA:ADMA201400192} and neutron depolarization imaging \cite{raey}. Furthermore, as the beamline is compatible to the whole sample environment available at FRM II, complex experiments involving non-ambient conditions (as high fields, low or high temperatures) can be performed. 

\subsection{The nGI setup}
\label{sec_nGI}
For the absorption grating G$_0$, a polished, single crystalline quartz wafer (diameter = \SI{102}{\milli\meter}, thickness = \SI{1}{\milli\meter}) was used as substrate. An adhesive layer of \SI{25}{\nano\meter} chromium followed by a neutron absorbing layer of \SI{20}{\micro\meter} gadolinium and a protective cover layer of \SI{50}{\nano\meter} aluminum was deposited onto the wafer by Ar sputtering. Gd has been chosen as absorbing material because it has the highest absorption cross section for thermal and cold neutrons \cite{doi:10.1080/10448639208218770}. The grating lines were subsequently incorporated into the layers by laser ablation. The resulting absorption grating has a periodicity of $p_{\mathrm{G0}}=\SI{1.6}{\milli\meter}$ and a duty cycle $DC_0$ of 0.4. The grating is mounted in a rotation stage (see Fig. \ref{Antares-Setup}), allowing for \SI{360}{\degree} rotation around the beam axis. Furthermore, the grating can be vertically inclined by an angle $\chi$ to change its effective periodicity $p_0 = p_{\mathrm{G0}}\cos(\chi)$ seen from the detector. The whole G$_0$ setup is mounted on an x-translation stage, allowing for high precision movement perpendicular to the neutron beam. 
\\The phase grating G$_1$ is made from Si, as Si reveals negligible attenuation for neutrons \cite{doi:10.1080/10448639208218770}. The gratings were commercially obtained from micromotive mikrotechnik \cite{Micromotive}. The rectangular grating lines were dry etched into the surface of the Si wafer (diameter = \SI{127}{\milli\meter}, thickness = \SI{0.5}{\milli\meter}). The resulting grating has a periodicity of $p_1 = \SI{7.98}{\micro\meter}$, a duty cycle of 0.5 and a structure height of $h_1=\SI{43}{\micro\meter}$.
\begin{figure}\centering%
\includegraphics[width=0.75\textwidth]{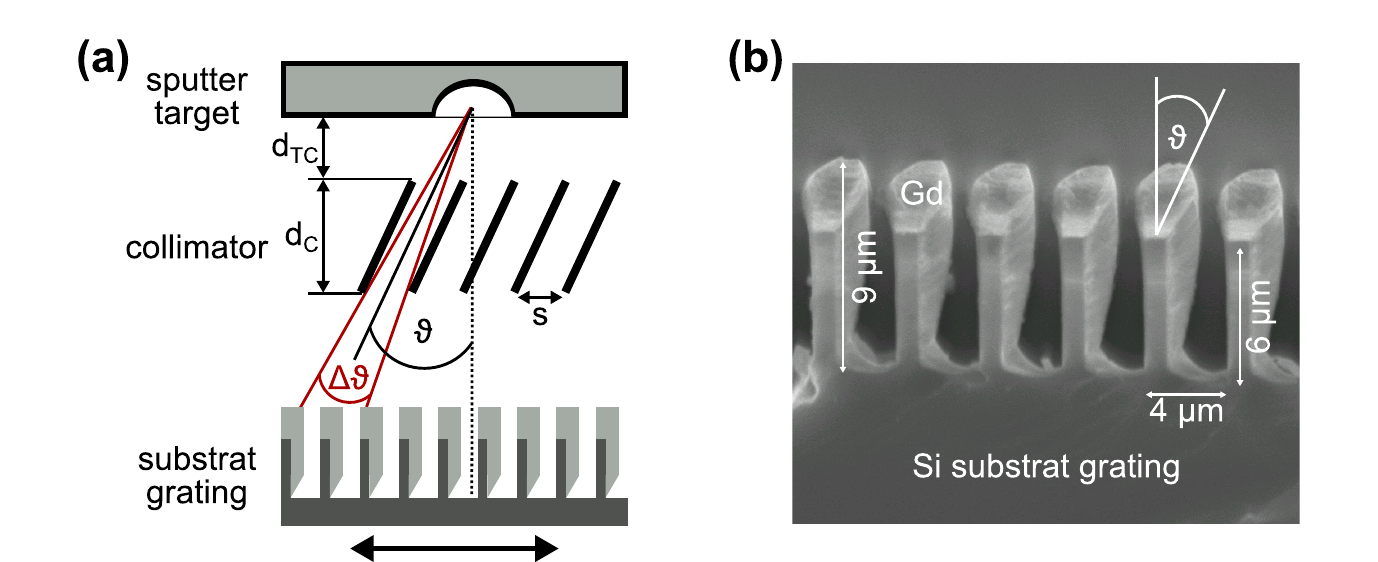}%
\caption{The fabrication of G$_2$ by Gd sputtering on a Si grating: a: Schematic of the sputtering geometry ($d_\mathrm{TC}=\SI{29}{\milli\meter}$, $d_\mathrm{C}=\SI{10}{\milli\meter}$ and $s=\SI{4.1}{\milli\meter}$)  b: SEM cross-sectional image of the obtained Gd absorption grating.}%
\label{Sputtering}%
\end{figure}
As the analyzer grating G$_2$ has the smallest periodicity, its fabrication is most challenging. So far, these gratings were produced following the method described in \cite{Grunzweig_Rev_Sci_Inst_2008}, in which Gd is sideways evaporated onto a Si grating having the required periodicity $p_2$. However, especially for large wafers, this procedure results in an inhomogeneous distribution of Gd on the substrate. Hence, we used a different approach based on Ar sputtering, which is illustrated in Fig. \ref{Sputtering} a: A collimator was introduced between the Gd sputtering target and the substrate grating that is continuously moved back and forth during the sputtering. The substrate grating has been structured on top of a \SI{127}{\milli\meter} Si wafer, similar to G$_1$, and has a periodicity of $p_2=\SI{4}{\micro\meter}$, a height of \SI{6}{\micro\meter}, and a grating line thickness of \SI{1}{\micro\meter}. The collimator consisted of thin brass lamellae inclined at $\vartheta=\SI{25}{\degree}$. In this geometry, only Gd atoms leaving the target at an angle of $\vartheta=\SI{25}{}\pm\SI{2.5}{\degree}$ can reach the substrate, allowing for a specific sidewall deposition. A scanning electron microscopy (SEM) image of the resulting grating is shown in Fig. \ref{Sputtering} b. The image was recorded on the cross-section of a small \SI{1x2}{\centi\meter^2} test piece of the grating covered with \SI{3}{\micro\meter} of Gd. A well defined absorption grating was obtained. The angle $\vartheta$ is recovered as the diagonal of the Gd deposition on top of the Si lines. The height of the Gd lines amounts $h_2=\SI{9}{\micro\meter}$ resulting in a maximal neutron transmission of \SI{7}{\percent} at $\lambda = \SI{3.5}{\angstrom}$. The sputtered G$_2$ revealed a total neutron transmission of \SI{59}{\percent} at $\lambda = \SI{3.5}{\angstrom}$ with an inhomogeneity of $\pm\SI{1.5}{\percent}$ over the whole grating. Hence, an effective duty cycle can be calculated to be $DC_2=\frac{\SI{59}{\percent}-\SI{7}{\percent}}{1-\SI{7}{\percent}}=\SI{0.56}{}$. The gratings G$_1$ and G$_2$ are mounted together on a large Euler cradle, which is situated in the second chamber directly in front of the detector (compare Fig. \ref{Antares-Setup}). This setup allows to rotate both gratings simultaneously around the beam axis. Furthermore, G$_1$ is fixed on a goniometer head and a linear stage, which enables to tune the distance $d$ of the gratings and to rotate G$_1$ against G$_2$, which is necessary to adjust the setup.

\section{Characterization of the setup}
\label{Sec_char}
In x-ray grating interferometry, especially at a synchrotron source, the setups are mostly optimized to exhibit a maximal visibility. This is guaranteed as long as the geometric relations of Sec. \ref{subdl} are fulfilled and $d$ corresponds to a fractional Talbot distance \cite{Grunzweig_Rev_Sci_Inst_2008}. However, neutron radiography suffers from the low neutron flux available, even at high brilliance neutron sources. Therefore, the setup optimization cannot be based on visibility considerations only, as a reduction of the setup length will strongly increase the neutron flux at the sample position. Based on these considerations, we reduced the setup length $L+d$ and moved the detector to the preferred sample position in chamber 2 (Fig. \ref{Antares-Setup}). The setup parameters were adjusted to comply with the relations derived in Sec. \ref{Sec_wavelength_V}. In Tab. \ref{Tab_parameter}, the actual parameters are shown and compared with the parameters calculated according to Ref. \cite{Grunzweig_Rev_Sci_Inst_2008}. By using Equ. \ref{End_visibility}, we can quantify the maximum visibility reduction caused by these improvements to only \SI{1.5}{\percent}. However, this is compensated by a flux enhancement of \SI{24}{\percent} at the detector, due to a reduced distance to the pinhole.
\begin{table}\centering
\begin{tabular}{|l|r|r|}
\hline
\textbf{parameter}& \textbf{value according to Ref. \cite{Grunzweig_Rev_Sci_Inst_2008}} & \textbf{actual value} \\ \hline
$\lambda$ & \SI{3.5}{\angstrom} & \SI{4.0}{\angstrom}\\ \hline
$p_0$ & \SI{1.596}{\milli\meter}& \SI{1.596}{\milli\meter}  \\ \hline
$DC_0$ &  \SI{0.4}{} & \SI{0.4}{}\\ \hline
$p_1$ & \SI{7.98}{\micro\meter} & \SI{7.98}{\micro\meter}\\ \hline
$p_2$ & \SI{4.00}{\micro\meter} & \SI{4.00}{\micro\meter} \\ \hline
$d$ & \SI{22.7}{\milli\meter} & \SI{18.0}{\milli\meter} \\ \hline
$L$ & \SI{9.10}{\meter} & \SI{7.19}{\meter} \\ \hline
\end{tabular}
\caption{Parameters of the nGI setup: $\lambda$ neutron wavelength, $p_{\mathrm{i}}$ periodicity of grating G$_{\mathrm{i}}$, $DC_0$ duty cycle of G$_0$, $L$ distance between G$_0$ and G$_1$ and $d$ distance from G$_1$ to G$_2$.}
\label{Tab_parameter}
\end{table}
In the following, the wavelength-dependence of the visibility of the nGI setup at ANTARES and the sensitivity of the DFI contrast for different particle sizes is characterized in detail. Furthermore, we will demonstrate the potential of quantitative DFI for structure analysis.

\subsection{Visibility of the setup}

\begin{figure}\centering%
\includegraphics[width=0.5\textwidth]{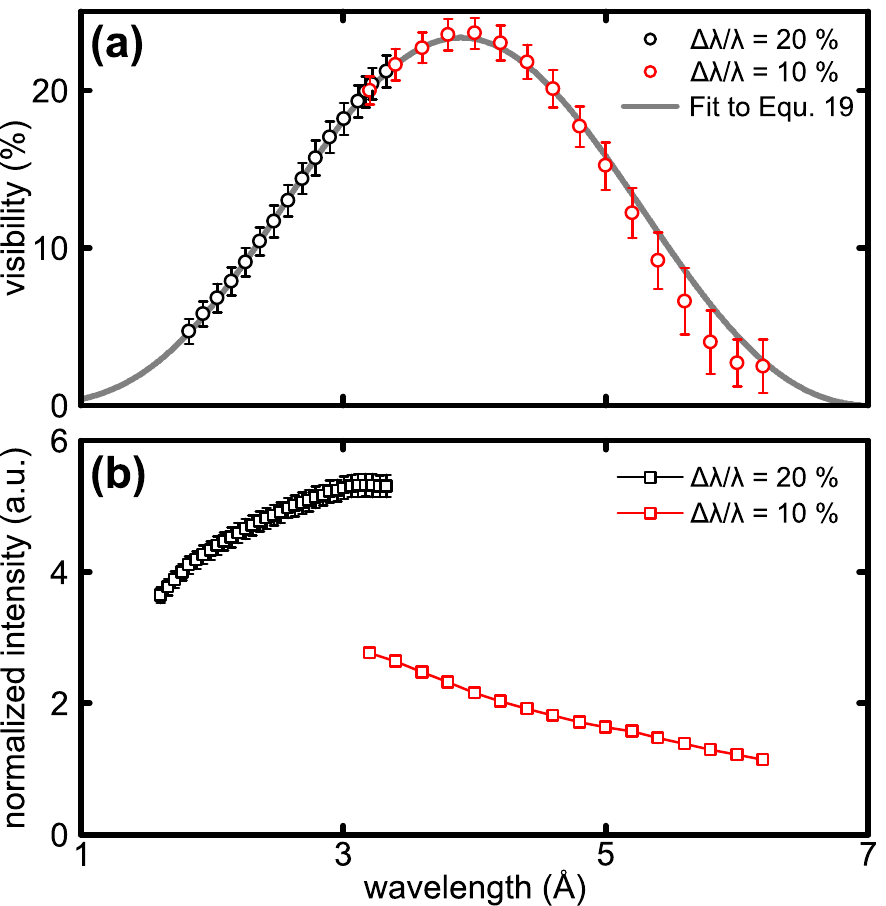}%
\caption{Characterization of the nGI: Visibility of the current setup (a) and normalized intensity at the detector (b) vs wavelength for two different $\Delta\lambda/\lambda$. The visibility was extracted by averaging the visibility map of the full field of view. Error bars correspond to the standard deviation of the visibility. The intensity was determined at a scintillator position next to the gratings and normalized to the exposure time. Error bars correspond to the statistical counting error.}%
\label{Visibility}%
\end{figure}
\begin{figure}\centering%
\includegraphics[width=0.5\textwidth]{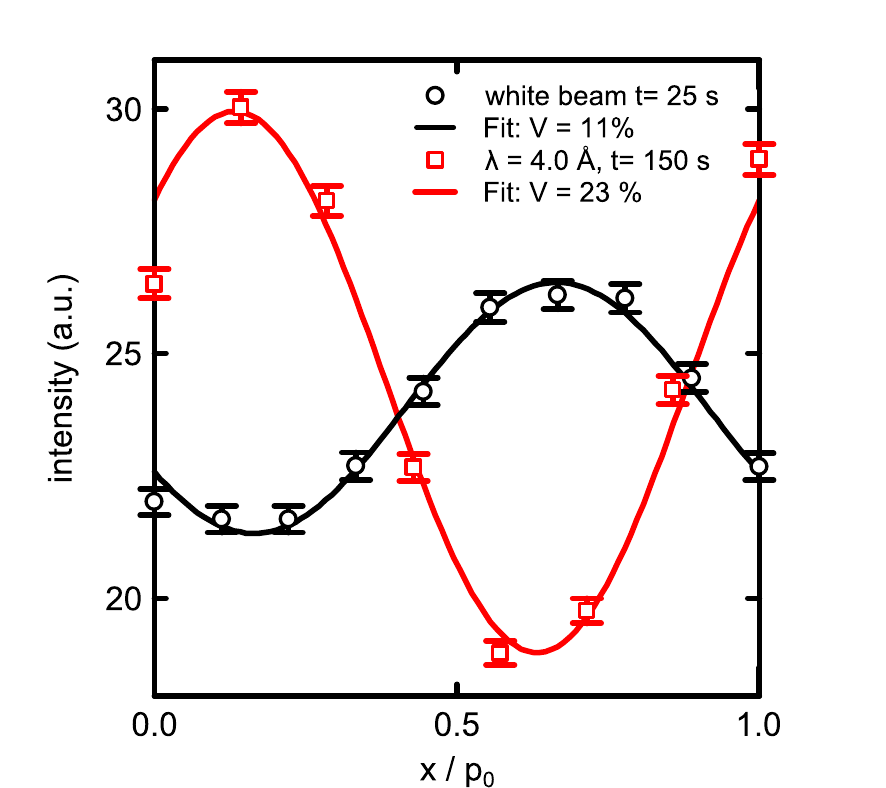}%
\caption{Monochromatic vs. white beam visibility: Intensity oscillation during an nGI stepping scan in a monochromatic and polychromatic beam. Error bars correspond to the statistical counting error. The exposure time per step was chosen to obtain a similar averaged intensity on the detector for both measurements.}%
\label{SteppingWB}%
\end{figure}

The wavelength dependence of the visibility of the setup is shown in Fig. \ref{Visibility} a. As the neutron velocity selector is unable to reach wavelengths below \SI{2.95}{\angstrom} in its default configuration (wavelength spread $\Delta\lambda/\lambda=0.1$), the NVS was tilted by \SI{5}{\degree} to access lower $\lambda$ \cite{FRIEDRICH1989547} in a second scan. However, this is accompanied by an increase of $\Delta\lambda/\lambda$ to \SI{0.2}{}. The visibility was determined for each pixel from a stepping scan over one period in 9 ($\Delta\lambda/\lambda=0.2$) or 8 steps ($\Delta\lambda/\lambda=0.1$), respectively and subsequently averaged over the whole image. The exposure time was set to \SI{80}{\second} per step. The maximum of the visibility is found between $\lambda=\SI{3.9}{\angstrom}$ and $\lambda=\SI{4.0}{\angstrom}$. The data were fitted using Equ. \ref{End_visibility} and the setup parameters defined in Table \ref{Tab_parameter} and Paragraph \ref{sec_nGI}. The maximum achievable visibility $V_0$ has been determined to be \SI{28.2}{\percent}. 
\\Obviously, the trend of the visibility is very well described by the derived relation (\ref{End_visibility}), although the wavelength was not purely monochromatic. However, an evaluation of Equ. \ref{NVS_distribution} reveals that the deviation from the monochromatic visibility $V$ is expected to be smaller than \SI{1}{\percent} for $\Delta\lambda/\lambda=0.1$ and \SI{1.7}{\percent} for $\Delta\lambda/\lambda=0.2$ in the particular wavelength range. Hence, the deviations caused by using the NVS lie well bellow the error bars obtained in Fig. \ref{Visibility}. Therefore, especially below $\lambda=\SI{3.3}{\angstrom}$ it is preferable to perform experiments with a higher wavelength spread, as the gain in the detected intensity is a factor of 2 in the case of increasing $\Delta\lambda/\lambda$ from \SI{10}{\percent} to \SI{20}{\percent} (see Fig. \ref{Visibility} b).
\\Measurements using the white beam (WB) of ANTARES may be necessary to obtain sufficient counting statistics for time-resolved measurements as well as in tomographic reconstructions \cite{Manke2010}. Introducing the ANTARES spectrum \cite{Tremsin2015} into Equ. \ref{Vis_distribution} reveals a visibility of \SI{10}{\percent}. To verify this value experimentally, Fig. \ref{SteppingWB} shows nGI data obtained in a monochromatic beam of \SI{4}{\angstrom} and in the WB. Presented is the intensity variation $I(x_{\mathrm{g}})$ in the center of the detector as function of the position of G$_0$ (stepping scan). To detect a similar averaged intensity, the exposure times per image were set to \SI{150}{\second} and \SI{25}{\second} for the mono- and polychromatic case, respectively. The WB visibility reduces to $\SI{11}{\percent}\pm\SI{1}{\percent}$, in agreement with the above prediction. However, as the exposure time is significantly reduced, white beam measurements provide a higher counting rate for the same exposure time and should be considered for all applications where no quantitative evaluation of the DPC and DFI signal is required. 

\subsection{The sensitivity of the DFI to structures of different sizes}
\label{QDFI}
To define the DFI sensitivity to different structure sizes, it has been proposed to use diluted spherical particles as reference material \cite{Lynch2011}. Although, the DFI contrast might be slightly different for arbitrarily shaped microstructures, this referencing has general significance for diluted systems. The reason is the property of the correlation function $G$ (Equ. \ref{DFI_xi}) to decay to zero at the longest distance characterizing the microstructure which is the diameter for spheres (see e.g. Ref. \cite{Andersson:aj5110}). The DFI sensitivity to different structure sizes can be calculated via Equ. \ref{DFI_xi}, as the correlation function $G$ and the macroscopic cross section $\Sigma$ are known as \cite{Andersson:aj5110}:
\begin{align}
\begin{split}
&G(\lambda,d_{\mathrm{col}})=\left(\left[1-\left(\frac{\xi_{\mathrm{GI}}(\lambda)}{d_{\mathrm{col}}}\right)^2\right]^{\frac{1}{2}}\left[1+\frac{1}{2}\left(\frac{\xi_{\mathrm{GI}}(\lambda)}{d_{\mathrm{col}}}\right)^2\right]\right. \\
&+\left. 2\left(\frac{\xi_{\mathrm{GI}}(\lambda)}{d_{\mathrm{col}}}\right)^2\left(1-\frac{\xi_{\mathrm{GI}}(\lambda)}{2d_{\mathrm{col}}}\right)^2\mathrm{ln}\left\{\frac{\frac{\xi_{\mathrm{GI}}(\lambda)}{d_{\mathrm{col}}}}{1+\left[1-\left(\frac{\xi_{\mathrm{GI}}(\lambda)}{d_{\mathrm{col}}}\right)^2\right]^{\frac{1}{2}}}\right\}\right)
\end{split}
\label{G_sphere}
\end{align}
and \cite{Strobl2014}
\begin{equation}
\Sigma=\frac{3}{4}\phi_{\mathrm{V}}\Delta\rho^2\lambda^2d_{\mathrm{col}},
\label{Sigma_sphere}
\end{equation}
respectively. Here, $d_{\mathrm{col}}$ is the diameter of the spheres, $\phi_{\mathrm{V}}$ is the particle concentration and $\Delta\rho$ the difference in neutron scattering length density of particle and solvent.
\\Following the discussed approach, nGI experiments were performed on different, di\-luted mono-dispersed polystyrene particles, similar to the ones used in Ref. \cite{Betz2015}. The spherical particles with a diameter of \SI{110}{\nano\meter}, \SI{510}{\nano\meter}, \SI{740}{\nano\meter}, \SI{1.0}{\micro\meter}, \SI{2.0}{\micro\meter}, \SI{3.0}{\micro\meter}, \SI{4.0}{\micro\meter} and \SI{5.0}{\micro\meter}, respectively, were dissolved in a mixture of \SI{56}{\percent} H$_2$O and \SI{44}{\percent} D$_2$O and each sample filled in a \SI{5}{\milli\meter} thick quartz cuvette. The particle volume concentration $\phi_{\mathrm{V}}$ was set to \SI{9}{\percent} for each particle diameter. $\Delta\rho$ has been calculated to be \SI{1.082d14}{\meter^{-2}} \cite{NIST}. nGI scans (16 steps, 1 period $p_0$, exposure time per step = \SI{60}{\second}) were performed on all of the colloides and on three cuvettes filled with H$_2$O, D$_2$O and the H$_2$O/D$_2$O solvent, respectively.
\begin{figure}\centering
\includegraphics[width=1\textwidth]{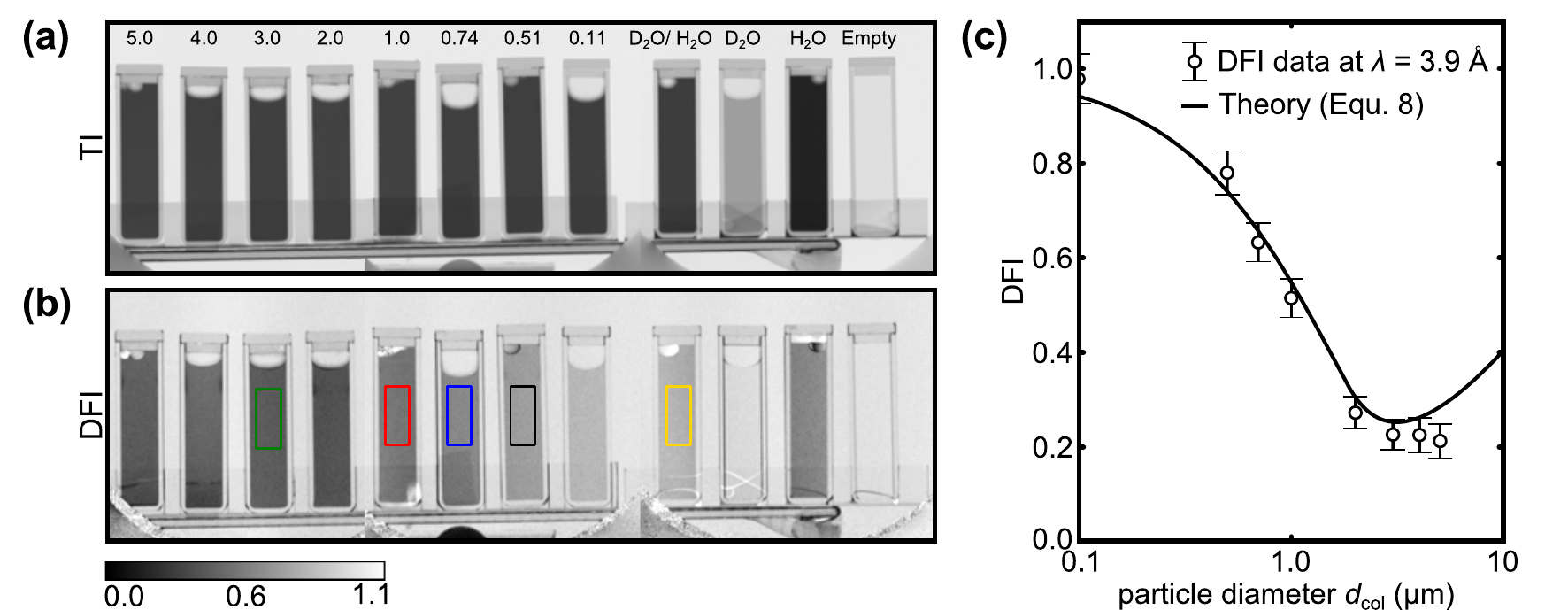}%
\caption{DFI sensitivity to different particle sizes. TI (a) and DFI (b) of cuvettes filled with diluted spherical colloids of different diameter, an empty cuvette and three cuvettes filled with  H$_2$O, D$_2$O and H$_2$O/D$_2$O, respectively. The images were merged from different TIs and DFIs recorded at $\lambda=\SI{3}{\angstrom}$. Clearly the DFI contrast depends on the diameter of the spheres, whereas the TI contrast is not influenced by the particle size. c: DFI versus particle diameter of the colloids at $\lambda=\SI{3.9}{\angstrom}$. The averaged DFI within each colloid was normalized to the contrast of the H$_2$O/D$_2$O mixture. The solid curve represents the theoretical sensitivity calculated for spherical particles. Error bars are calculated by error propagation from the DFI standard deviation in the probed areas.}%
\label{quantitativeDFI}%
\end{figure}
Fig. \ref{quantitativeDFI} a and b show the TIs and DFIs of all the samples, respectively, taken at a wavelength of \SI{3}{\angstrom}. The images were stitched from three TIs (DFIs), recorded separately. The TI contrast is nearly identical for all colloids, as their chemical composition and concentration is equal. In contrast, the DFIs shown in Fig. \ref{quantitativeDFI} b reveal strong deviations in the signal for the different colloids: the high contrast for the \SI{0.1}{\micro\meter} particles strongly decreases with enhanced particle sizes. To quantify the contrast degradation, Fig. \ref{quantitativeDFI} c shows the average DFI of the colloids as function of the particle diameter $d_{\mathrm{col}}$ for a neutron wavelength of \SI{3.9}{\angstrom}, corresponding to the value generating maximal visibility (see Sec. \ref{Sec_wavelength_V}). To eliminate the DFI contrast contribution, arising from incoherent scattering at the hydrogen within the solvent \cite{Betz2015}, the DFI values were normalized to the DFI within the H$_2$O/D$_2$O mixture taken to the power of \SI{91}{\percent}. This fractional normalization is slightly different to the approach in Ref. \cite{Betz2015} and accounts for the fact that in the colloids \SI{9}{\percent} of the solvent is replaced by polystyrene\footnote{As the effective thickness of the incoherently scattering solvent is reduced by \SI{10}{\percent}. However, the correct normalization of DFI data is still a question of debate and demands further investigations.}.
\\The DFI contrast clearly decreases from nearly unity at a particle diameter of \SI{0.11}{\micro\meter} towards 0.2 at \SI{3}{\micro\meter}, in agreement with the theory curve calculated by introducing Equations \ref{G_sphere} and \ref{Sigma_sphere} into \ref{DFI_xi}. The expected increase of the DFI for large $d_\mathrm{col}$ cannot be deduced from the presented data of colloids with $d_\mathrm{col}<\SI{5}{\micro\meter}$. However, the increase has been observed for larger particles \cite{Betz2015}. Nonetheless, the data clearly confirm the statements above, that the DFI is insensitive to structures smaller than \SI{0.1}{\micro\meter} and is mostly sensitive to particles in the micron range generating USANS scattering.

\subsection{The potential of quantitative DFI}
\label{potential_QDFI}
\begin{figure}\centering
\includegraphics[width=0.5\textwidth]{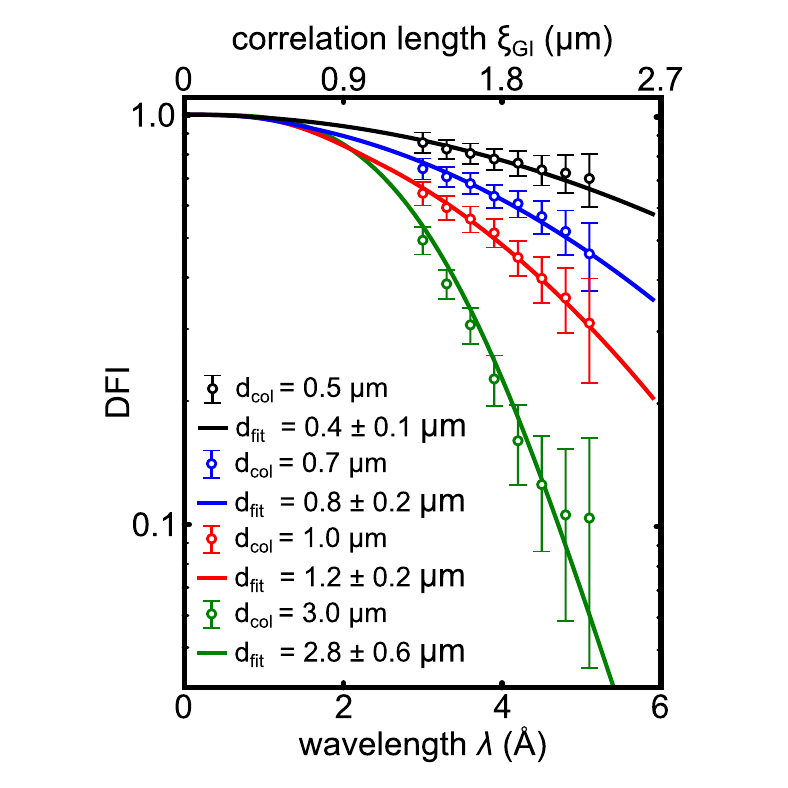}%
\caption{Quantitative evaluation of the DFIs vs. wavelength. The DFI signal was averaged within the regions marked in Fig. \ref{quantitativeDFI} b, normalized to the DFI contrast of the H$_2$O/D$_2$O mixture (yellow box) and plotted against the wavelength. The data were fitted according to Equ. \ref{G_sphere} and \ref{Sigma_sphere}. The $d_{\mathrm{fit}}$ values are given in the legend. Error bars are calculated by error propagation from the DFI standard deviation in the probed areas.}%
\label{Fig_DFI_wavelength}%
\end{figure}

The presented DFI sensitivity to structures of different sizes (Sec. \ref{QDFI}) can be directly applied for the quantification of micrometer sized precipitations or small pores within the objects to be investigated. For more complex or dense microstructures, the potential of a single nGI scan to obtain quantitative information is limited. Nonetheless, by a variation of $\xi_{\mathrm{GI}}$ via a $\lambda$- or $L_{\mathrm{S}}^{\mathrm{eff}}$- scan the correlation function $G$ can be probed over a broader length scale. The accuracy of this approach is demonstrated in Fig. \ref{Fig_DFI_wavelength} which shows the wavelength dependence of the DFI signal for colloids with particle diameter of \SI{0.51}{\micro\meter}, \SI{0.74}{\micro\meter}, \SI{1.0}{\micro\meter} and
\SI{3.0}{\micro\meter}, respectively. The DFIs were normalized to the solvent as discussed above. The evaluated regions within the cuvettes are marked in Fig. \ref{quantitativeDFI} b. Using Equ. \ref{G_sphere} and \ref{Sigma_sphere} the DFI data was fitted and the particle diameters of the colloids $d_{\mathrm{fit}}$ were determined. 
The presented model describes the contrast variation well and provides a good estimate of the particle diameters. However, the size determination is based on the a priori knowledge of the underlying structure (shape, concentration, chemical composition). A general structure determination is limited in grating based methods, by the limitations on the probed correlation length $\xi_{\mathrm{GI}}$. Nonetheless, the quantitative DFI approach provides spatially resolved information which are strongly complementary to results from scattering methods such as SANS and USANS. Hence, structural information obtained by means of these scattering techniques can be used to determine an averaged correlation function \cite{Andersson:aj5110} of the microstructure which can then be checked by means of nGI for local deviations in e.g. shape, concentration or structure size. Therefore, in combination with scattering techniques the quantitative DFI approach may have significance for the investigation of e.g. domain nucleation in arbitrary systems as in e.g. ferromagnets or superconductors \cite{Reimann2015a}, particle sedimentation or phase precipitations.

\section{Typical Applications of nGI}
\label{Sec_app_nGI}
\subsection{Material differentiation and testing using the DFI contrast modality}
\label{Mat_diff}
\begin{figure}\centering%
\includegraphics[width=1\textwidth]{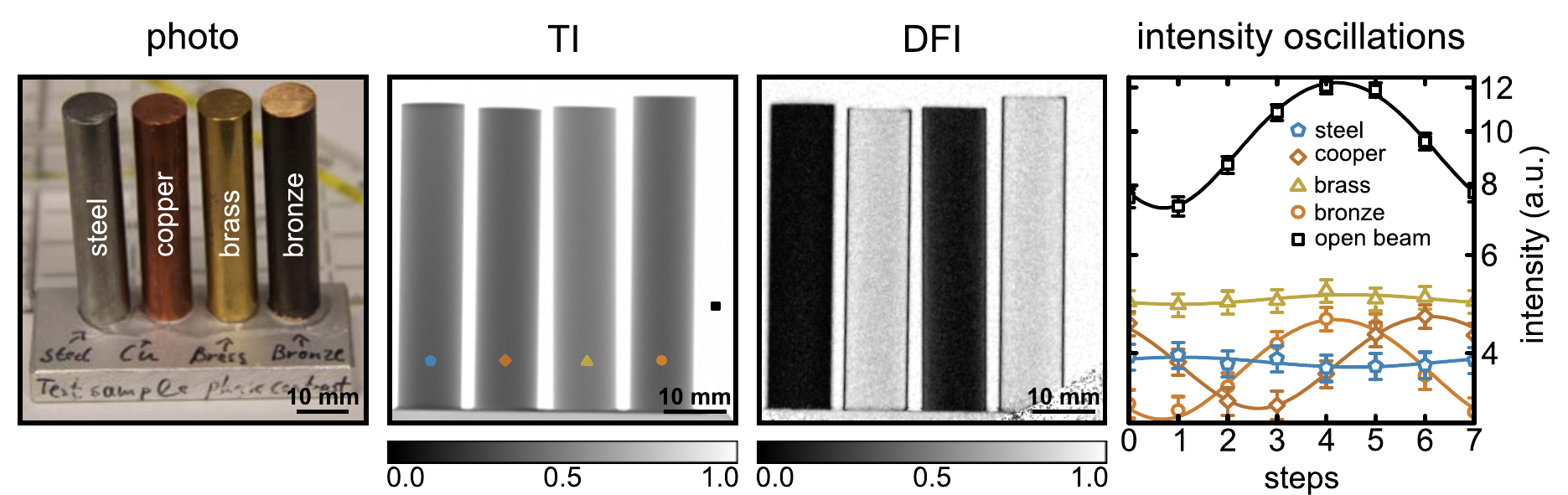}%
\caption{nGI for material differentiation: Photo, transmission image (TI) and dark-field image (DFI) of a test sample consisting of a steel, a copper, a brass and a bronze rod. The wavelength was set to $\lambda=\SI{4.0}{\angstrom}$. In addition, the intensity oscillation during a stepping scan is shown for the pixels marked in the TI. The neutron absorption and hence the TI contrast of the materials is similar, whereas the DFI reveals a strong contrast degradation for steel and brass.}%
\label{Testsample}%
\end{figure}

The ability of nGI to differentiate materials showing similar neutron transmission, but different microstructural properties is demonstrated in the following. Fig. \ref{Testsample} shows nGI data of a test object composed of rods of steel, copper, brass and bronze, each having a diameter of \SI{10}{\milli\meter}. The dataset consists of a photograph, a neutron TI and a neutron DFI of the rods. Furthermore, the intensity oscillations $I(x_{\mathrm{g}})$ during a stepping scan are shown for the pixels marked in the TI. The TI and DFI were calculated from a scan of G$_0$ over 1 complete period $p_0$ in 8 equidistant steps. The exposure time per step was \SI{80}{\second}. 
Obviously, the TI which corresponds to the average of the shown oscillations, is similar for all materials with the exception of the brass rod, revealing a slightly higher transmission. On the other hand, the DFI signal, given by the amplitude, reveals a strong contrast degradation for steel and brass, which is attributed to scattering at magnetic domain walls \cite{PhysRevLett.101.025504} and at small precipitations within the material \cite{PhysRevB.88.125104}, respectively. In contrast, the DFI signal in the copper rod and its bronze alloy is less influenced. While this is anticipated for the pure and homogeneous metal Cu, a lower DFI contrast might be expected for the bronze, as it is composed of different chemical phases. However, the length scale of the segregations of these phases in bronze or their scattering contrast does not match the maximum sensitivity length of the DFI. The above example demonstrates the complementarity of TI and DFI for neutron imaging. However, the chosen materials demonstrate also the magnetic sensitivity of nGI \cite{Manke2010}\cite{PhysRevLett.101.025504}\cite{:/content/aip/journal/apl/93/11/10.1063/1.2975848} and the possibility to mark $\upmu$m structures as precipitations in alloys or porosities and cracks in cast materials \cite{:/content/aip/journal/jap/107/3/10.1063/1.3298440}. Both may have technical relevance in e.g. engineering and material science.

\subsection{Identification of micro textures and anisotropies}
\label{Directional_DFI}
Directional dark-field imaging evaluates the variation of the DFI signal with the rotation angle $\omega$ of the grating lines around the beam axis. The DFI is insensitive to the scattering components parallel to the grating lines. Hence, this rotation of the gratings will result in an oscillation of the DFI contrast if the microstructure of the sample is anisotropic. From the shape of the oscillation, detailed information about the microstructural orientation within a sample can be obtained. Depending on the scattering strength and the number of predominant scattering directions within the sample, different evaluation procedures have been published \cite{PhysRevB.82.214103}\cite{:/content/aip/journal/jap/112/11/10.1063/1.4768525}\cite{PhysRevB.84.094106}. In the following example we will assume a uniaxial orientation of the microstructure as was done in Ref. \cite{PhysRevB.82.214103}. In this case, the scattering function $S(\mathbf{q})$ can be approximated by an anisotropic 2D Gaussian and the $\omega$-variation of the DFI in each pixel (j,l) is given by:
\begin{equation}
DFI(\omega,j,l)=b_0(j,l)\exp\left\{-b_1(j,l)\sin^2\left(\omega-\Psi(j,l)\right)\right\} 
\label{DFI_omega}
\end{equation}
The coefficients $b_0$, $b_1$ and $\Psi$ specify the isotropic, $\omega$-independent DFI contribution, the aniso\-tropy of the DFI($\omega$) and the grating direction revealing the highest DFI contrast, respectively.
\begin{figure}\centering%
\includegraphics[width=1\textwidth]{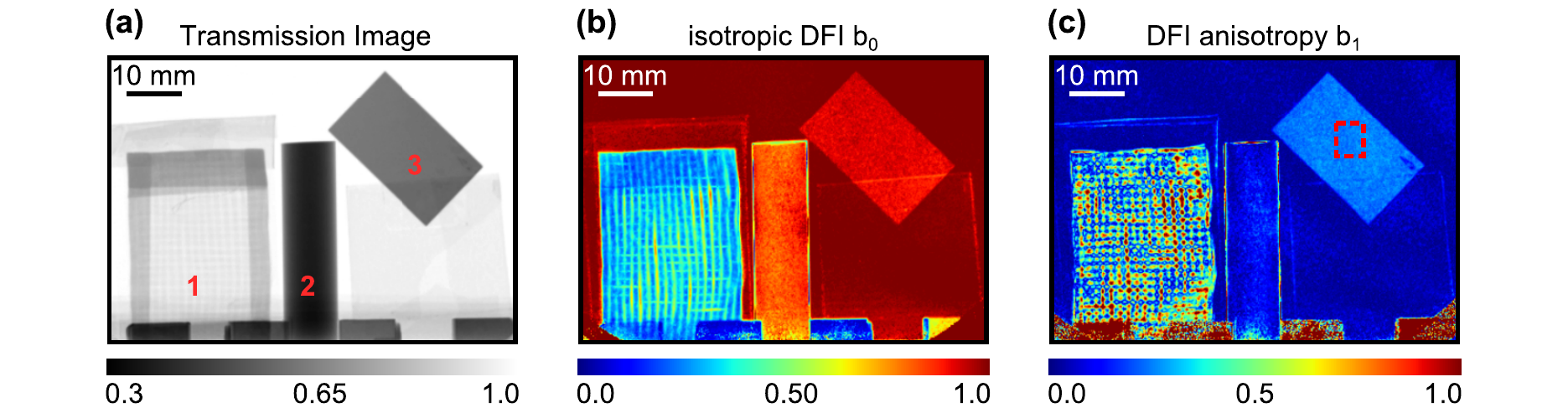}%
\caption{Directional dark-field imaging of a glass fiber mat (1), a copper rod (2) and a \SI{4}{\micro\meter} Gd grating (3). The figure shows a TI of the test samples (a), the isotropic part of the DFI contrast $b_0$ (b) and the DFI anisotropy $b_1$ (c). Details of the image reconstruction are given in the text.}%
\label{Directional-DFI}%
\end{figure}
The potential of directional DFI at ANTARES is demonstrated in Fig. \ref{Directional-DFI}. A woven glass fiber mat (1), a copper rod (2) and a test piece of a grating G$_2$ (3) were used as samples. DFIs of the objects were recorded at 10 different angular positions $\omega$ between \SI{-48}{\degree} and \SI{42}{\degree} with respect to the vertical alignment of the gratings. Each DFI was calculated from a stepping sequence of 15 images taken at a wavelength of \SI{4.0}{\angstrom}. The exposure time was set to \SI{150}{\second} per step. The parameters $b_0$ and $b_1$ were extracted for each pixel, by fitting the obtained DFIs to Equ. \ref{DFI_omega}. The resolution was approximately \SI{0.4}{\milli\meter}. 
\\The results shown in Fig. \ref{Directional-DFI} are divided into the transmission image (a), maps of the isotropic DFI $b_0$  (b) and the DFI anisotropy $b_1$  (c). Additionally, Fig. \ref{Directional-DFI-polar} (a) shows a photo of the weaving pattern of the mat partially overlaid by its isotropic DFI and the DFI anisotropy at the corresponding positions.
\\The isotropic DFI contrast is lowest in the glass fiber mat (1) precisely on a squared lattice where bundles of vertical and horizontal fibers overlay and the thickness of the mat is highest. The minima are surrounded by lines of higher isotropic DFI contrast which correspond to the lines in between the ribbons. These positions are schematically sketched as green lines in Fig. \ref{Directional-DFI-polar} a and b, which shows an illustration of the weaving pattern. The $b_0$ line pattern matches the position of maximal TI, which supports this interpretation. A similar square pattern is observed in the DFI anisotropy map of the mat (Fig. \ref{Directional-DFI} c) marking the points of highest scattering anisotropy. The maxima correspond to positions within the weaving pattern, in which mostly a single oriented bundle is in the beam (red squares in Fig. \ref{Directional-DFI-polar} b). In contrast, the minima correspond to positions where either perpendicular bundles overlay and their anisotropy cancel, or positions in which no ribbon is in the beam (blue circles). 
\begin{figure}\centering%
\includegraphics[width=0.4\textwidth]{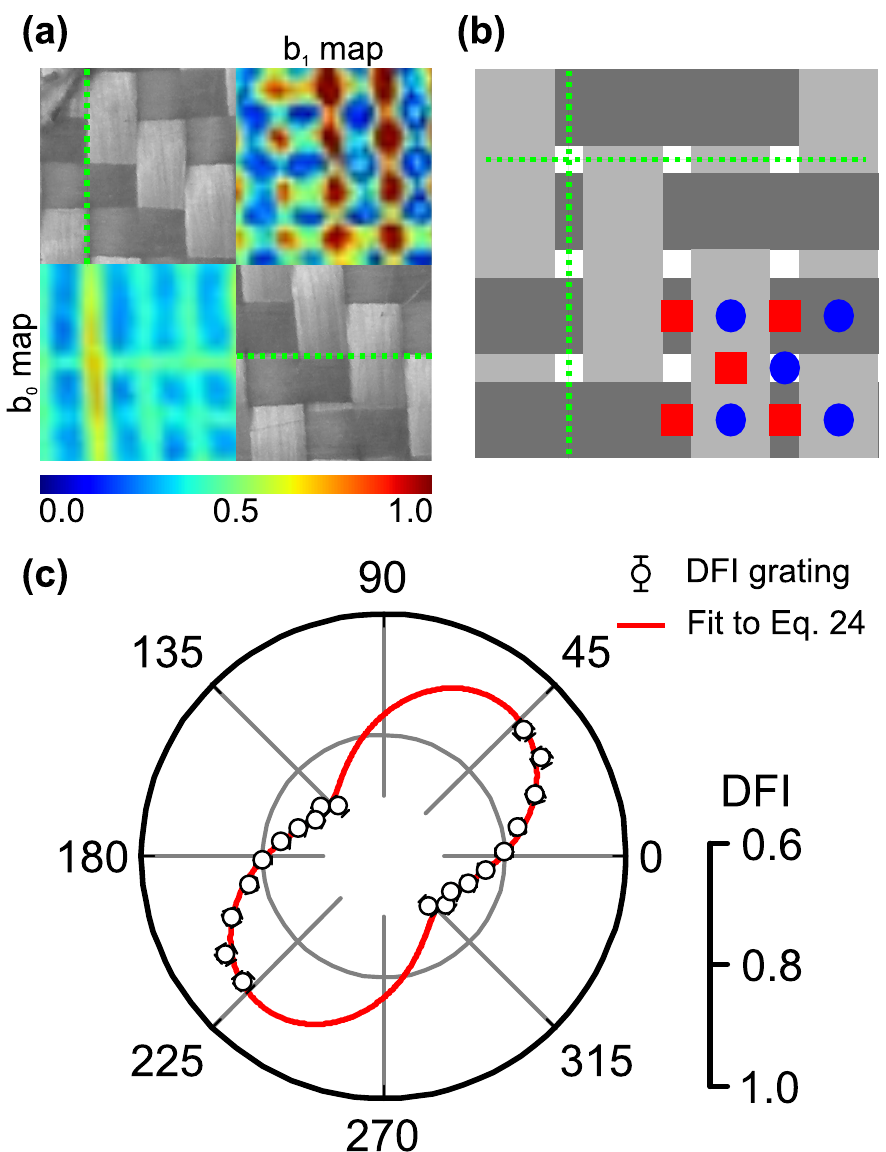}%
\caption{Interpretation of directional dark-field imaging. a: Zoomed photo of the glass fiber mat partially overlay by its isotropic DFI (left) and the DFI anisotropy at the corresponding positions. b: Illustration of the weaving pattern. Sketched are the lines of lowest absorption and scattering (green dotted) as well as the points of highest (red squares) and lowest anisotropy (blue circles). c: Polar plot of the DFI vs the grating rotation angle for the area marked in Fig. \ref{Directional-DFI} c. To extract the alignment of the test grating (3), the data were fitted to Equ. \ref{DFI_omega}. As the DFI has \SI{180}{\degree} symmetry in principle, the DFI data were additionally drawn at $\omega+\SI{180}{\degree}$. However, this does not influence the fit. Error bars correspond to the standard deviation of the DFI in the probed area.}%
\label{Directional-DFI-polar}%
\end{figure}
The situation differs in the Cu rod (2). Here the DFI contrast is isotropic as expected for a pure metal. Hence, $b_0$ follows the cylindrical shape of the sample, whereas the anisotropy map is nearly zero, apart from an edge enhancement due to refraction. The isotropic DFI is highest in the grating (3) and is generated by scattering within the layer of \SI{3}{\micro\meter} Gd which has been deposited onto the Si substrate grating. In contrast, the anisotropic part originates from the directed scattering off the grating structure. By plotting the mean DFI values of the region marked in (c) versus the rotation angle of the setup $\omega$, shown in Fig. \ref{Directional-DFI-polar} c, even the alignment of the grating lines can be extracted. For this, the data were fitted using Equ. \ref{DFI_omega}. The maximum of the DFI determined at \SI{51}{\degree} precisely corresponds to the configuration in which the setup is rotated perpendicular to the grating lines of the test piece. 
The presented directional DFI results show the ability of nGI to detect and quantify anisotropies within the microstructure of the material and to extract the preferred scattering directions. Hence, the method can be used to study the development of structural or magnetic anisotropies on the $\upmu$m-scale in situ during e.g. temperature variations and external stresses. Furthermore, directional DFI can have high technical relevance for the mapping of oriented structures within modern composite materials.

\section{Conclusion}

In conclusion, we presented the technical details of the newly implemented neutron grating interferometer at the ANTARES beamline. Moreover, a simple model has been developed to optimize the visibility and flux of an nGI setup according to the requirements of the experiments. Because of the high neutron flux at ANTARES and the simple implementation of additional spectrum shaping devices and sample environments (e.g. cryostats, magnets and furnaces) to the instrument, this nGI is one of the most flexible setups available. Furthermore, we demonstrated and discussed its potential on selected case studies: dark-field imaging for materials differentiation, identification of $\upmu$m anisotropies by directional dark-field imaging and the extraction of microstructural size information (i.e. the autocorrelation function) by means of quantitative DFI. 
\\The presented examples are thoroughly based on the particular DFI contrast mechanism, which is influenced by USANS scattering within the sample. Hence, nGI provides spatially resolved information about the samples microstructure within the bulk. By this means nGI closes the gap between the reciprocal scattering techniques as SANS and USANS and real space radiography. Therefore, the setup allows various novel experiments in fields as e.g. (magnetic) domain studies and material science but can be of industrial interest for e.g. material testing and characterization as well. The nGI setup is available for external user \cite{MLZ-Prop}.


\appendix
\ack{T.R. likes to thank Dominik Bausenwein, Tobias Neuwirth, Wolfgang Kreuzpaintner, as well as Michael Schneider (SwissNeutronics) for technical support and the Crystal and Material Laboratory of the TUM for the sample preparation. We highly appreciate the fruitful discussions with Ralph Harti and Christian Gr\"unzweig. Furthermore, we like to offer our special thanks to Benedikt Betz for providing the data used for the quantitative dark-field evaluation. This project has received funding from the European Union's Seventh Framework Programme for research, technological development and demonstration under the NMI3-II Grant number 283883.}

\referencelist[20160222_Bibliography_nGI]
\end{document}